\documentclass{ws-procs9x6}

\begin{document}

\title{First Results from a Multi-wavelength Survey of Quasar Jets}

\author{J.~M. GELBORD}

\address{MIT Center for Space Research, NE80-6091\\
77 Massachusetts Ave. \\ 
Cambridge, MA 02139, USA\\ 
E-mail: jonathan@space.mit.edu}

\author{H.L. Marshall, D.A. Schwartz, D.M. Worrall, M. Birkinshaw,
  J. Lovell, D. Jauncey, E. Perlman, D. Murphy and R.A. Preston}


\maketitle

\abstracts{
We present results based on the first 20 \textsl{Chandra} images
obtained in a survey of jets in radio selected flat-spectrum quasars
(FSRQs),
along with new sub-arcsecond radio maps and optical images. 
We discover jet X-ray flux in 12 sources despite short exposures,
establishing that FSRQ jets are often X-ray bright.
The X-ray morphology typically 
matches the radio and fades
rapidly after the first sharp radio
bend, but there are notable exceptions.
Optical non-detections rule out simple synchrotron models for jet
X-ray emission, 
implying these systems are dominated by inverse Compton
(IC) scattering. 
Models of IC scattering of the cosmic microwave background (CMB)
%
%
constrain the bulk flow and magnetic field, 
suggesting the jets are oriented close to our line
of sight, 
with deprojected lengths often $\gg 100$~kpc.
}

We are conducting a survey of a large, flux-limited sample of FSRQs
selected by extended 5~GHz flux ($> \! 2''$ from the core).
The 56 sample members span a wide range of radio morphologies and
redshifts.
Of these, 20 were observed by \textsl{Chandra} during
cycle~3\cite{marshall04}.
New sub-arcsecond resolution radio maps for all 20 were obtained with
\textsl{ATCA} and \textsl{VLA}, and six (so far) have been imaged with
\textsl{Magellan}\cite{gelbord04}.
Results are summarized in Table~\ref{table1}.

\begin{table}[htp]
\tbl{Some results for the 20 targets observed during \textsl{Chandra}
  cycle~3. \label{table1}}
{\footnotesize
\begin{tabular}{@{\extracolsep{0.05em}}@{}lccccccc@{}}
\hline
Target name   &   z   &  Bj mag  & A/B$^a$
                                     &X-jet?$^b$
                                          &$\alpha_{\mathrm{rx,jet}}$
                             & $g_{\mathrm{pred}}^c$ & $g_{\mathrm{obs}}^d$\\
\hline
\multicolumn{8}{c}{}\\[-2ex]
PKS 0208--512 & 0.999 & 17.1 &  B  &  Y  &$0.92\pm.01$&  23.4 &$>\!24.5$\\
PKS 0229+131  & 2.059 & 17.7 &  B  &  N  &  $>0.95$   &$\cdots$&$\cdots$ \\
PKS 0413--210 & 0.808 & 18.7 & A/B &  Y  &$1.04\pm.02$&  23.2 &$\cdots$ \\
PKS 0745+241  & 0.410 & 19.0 &  B  &  N  &  $>0.91$   &$\cdots$&$\cdots$ \\
PKS 0858--771 & 0.490 & 17.9 &  B  &  N  &  $>0.99$   &$\cdots$&$>\!26.0$\\
PKS 0903--573 & 0.695 & 19.0 & A/B &  Y  &$1.07\pm.02$&  22.9 &$\cdots$ \\
PKS 0920--397 & 0.591 & 18.3 &  A  &  Y  &$1.00\pm.02$&  23.4 &$\cdots$ \\
PKS 1030--357 & 1.455 & 19.5 &  B  &  Y  &$0.93\pm.01$&  23.0 &$>\!26.0$\\
PKS 1046--409 & 0.620 & 17.5 & A/B &  Y  &$0.95\pm.03$&  24.5 &$>\!25.5$\\
PKS 1145--676 & \phantom{$^e$}?$^e$ 
                      & 19.4 &  B  &  N  &  $>0.95$   &$\cdots$&$\cdots$ \\
PKS 1202--262 & 0.789 & 19.8 & A/B &  Y  &$0.86\pm.01$&  22.5 &$>\!26.7$\\
PKS 1258--321 & 0.017 & 13.0 &  A  &  Y  &$1.03\pm.03$&  23.7 &$\cdots$ \\
PKS 1343--601 & 0.013 & \phantom{$^f$}10.8$^f$
                             & A/B &  Y  &$1.01\pm.02$&  23.3 &$\cdots$ \\
PKS 1424--418 & 1.522 & 18.0 &  B  &  N  &  $>0.91$   &$\cdots$&$\cdots$ \\
PKS 1655+077  & 0.621 & 18.8 &  B  &  N  &  $>0.88$   &$\cdots$&$\cdots$ \\
PKS 1655--776 & 0.094 & 17.5 &  A  &  N  &  $>1.07$   &$\cdots$&$\cdots$ \\
TXS 1828+487  & 0.692 & 17.1 &  A  &  Y  &$0.91\pm.01$&  21.9 &$\cdots$ \\
PKS 2052--474 & 1.489 & 18.2 &  B  &  N  &  $>0.89$   &$\cdots$&$\cdots$ \\
PKS 2101--490 & \phantom{$^e$}?$^e$ 
                      & 17.1 &  B  &  Y  &$0.99\pm.02$&  23.8 &$>\!25.3$\\
PKS 2251+158  & 0.859 & 16.6 & A/B &  Y  &$0.95\pm.01$&  22.4 &$\cdots$ \\[0.5ex]
\hline
\end{tabular}}
\begin{tabnote}
$^a$ Membership in extended 5~GHz radio flux-limited (A)
     and/or morphologically-selected (B) subsamples.
$^b$ Is jet detected in X-rays?
$^c$ $g'$ mag predicted by synchrotron model (using $\alpha_\mathrm{rx,jet}$).
$^d$ Observed jet limiting mag (no optical detections).
$^e$ z to be measured with \textsl{Magellan} in 2004.
$^f$ Gunn $z$ mag; Bj extinction $\sim \! 12$ mag!
\end{tabnote}
\end{table}

\begin{figure}[htp]
\begin{minipage}[l]{0.1875\textwidth}
\centering
\includegraphics[height=1.23in]{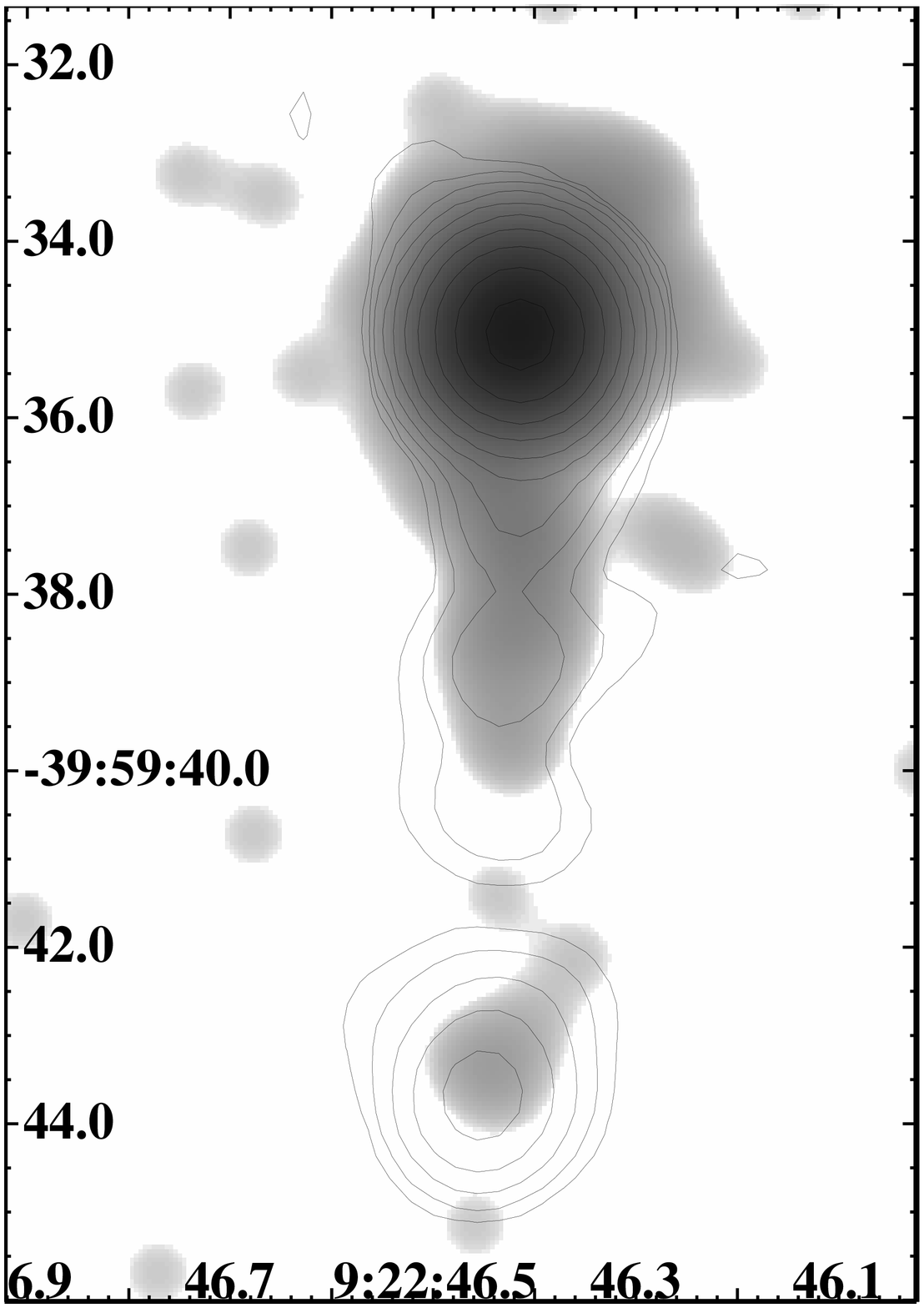}
\end{minipage}
\begin{minipage}[l]{0.1875\textwidth}
\centering
\includegraphics[height=1.23in]{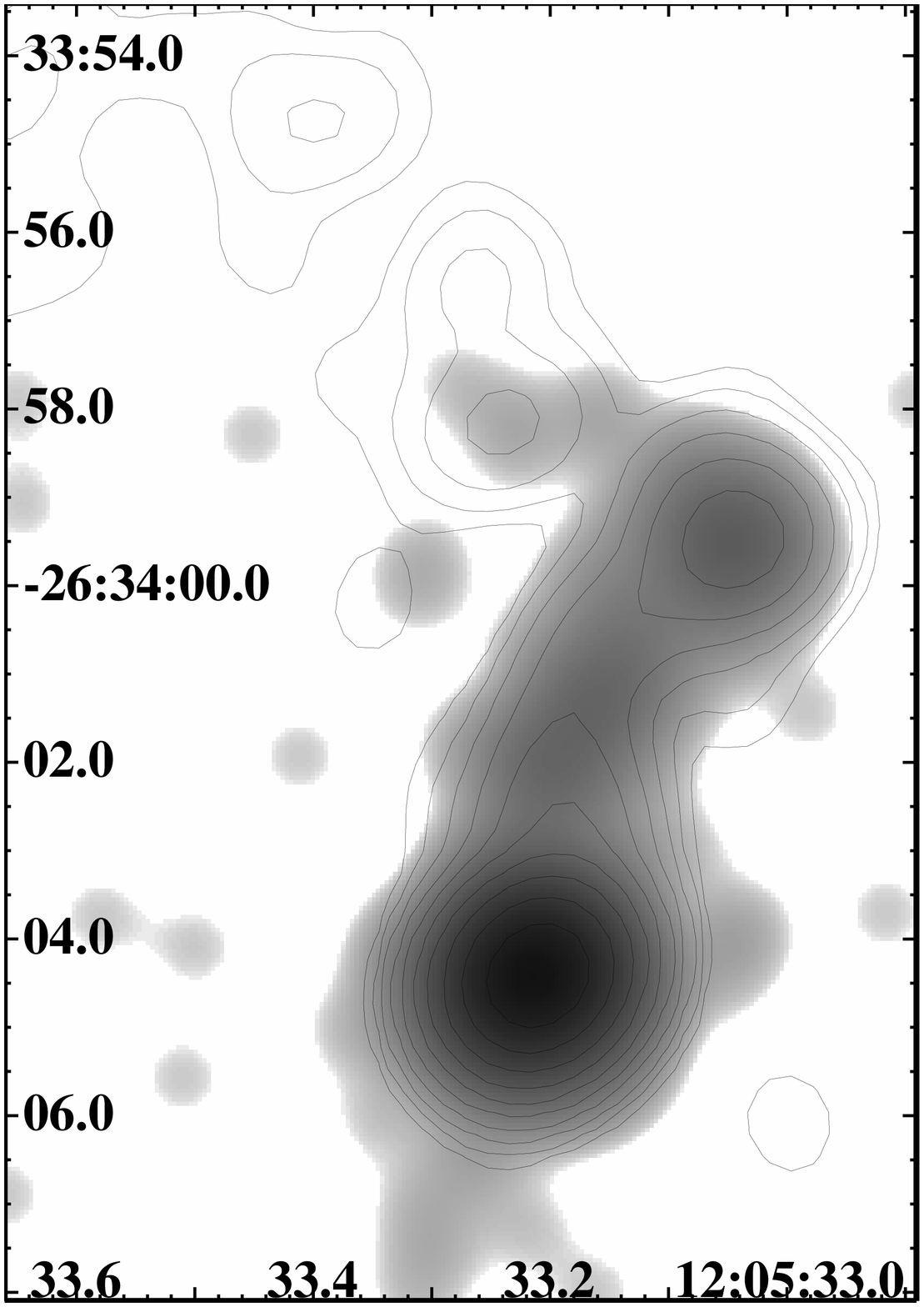}
\end{minipage}
\begin{minipage}[l]{0.24\textwidth}
\centering
\includegraphics[height=1.23in]{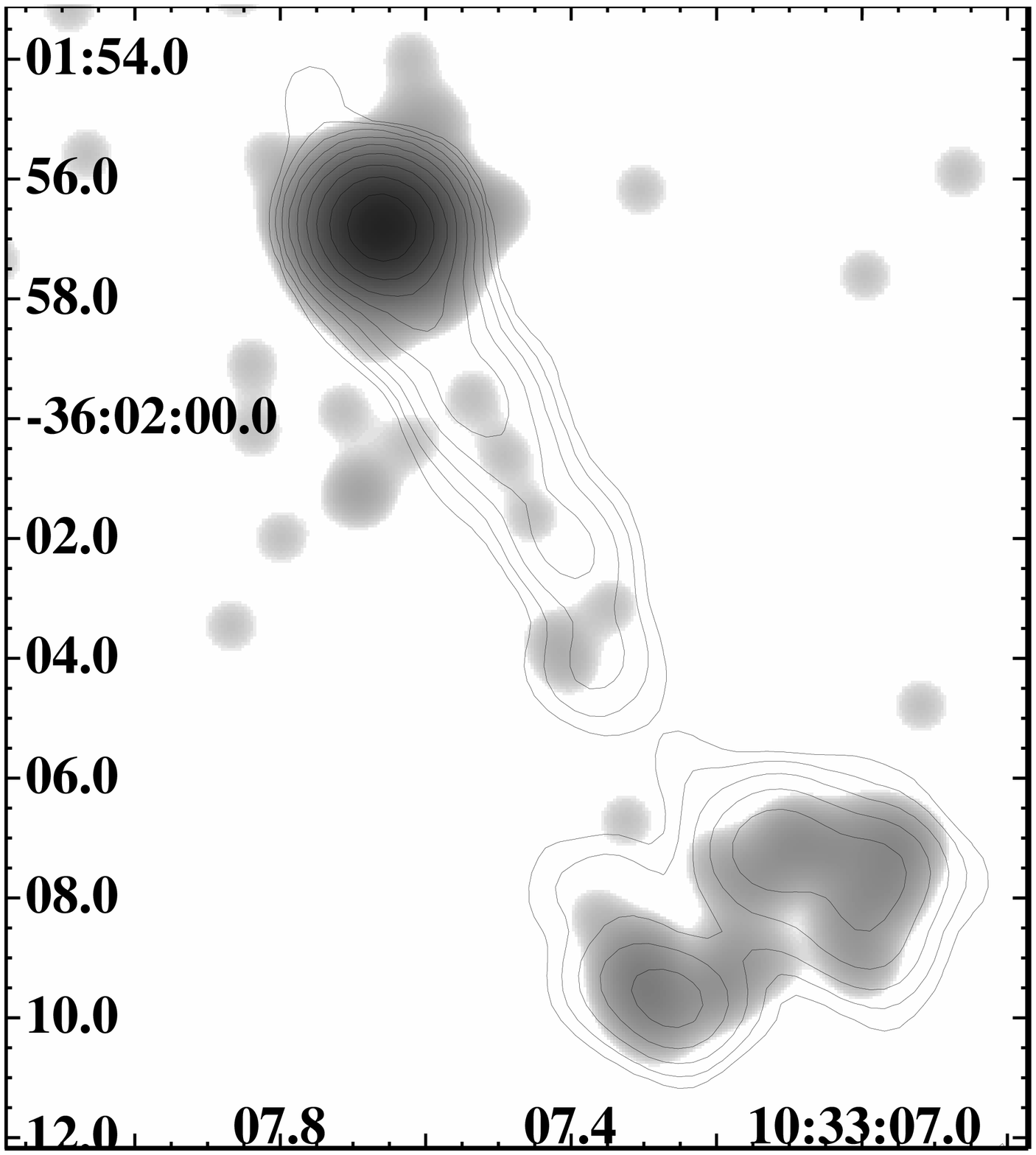}
\end{minipage}
\begin{minipage}[l]{0.34\textwidth}
\centering
\includegraphics[height=0.82in]{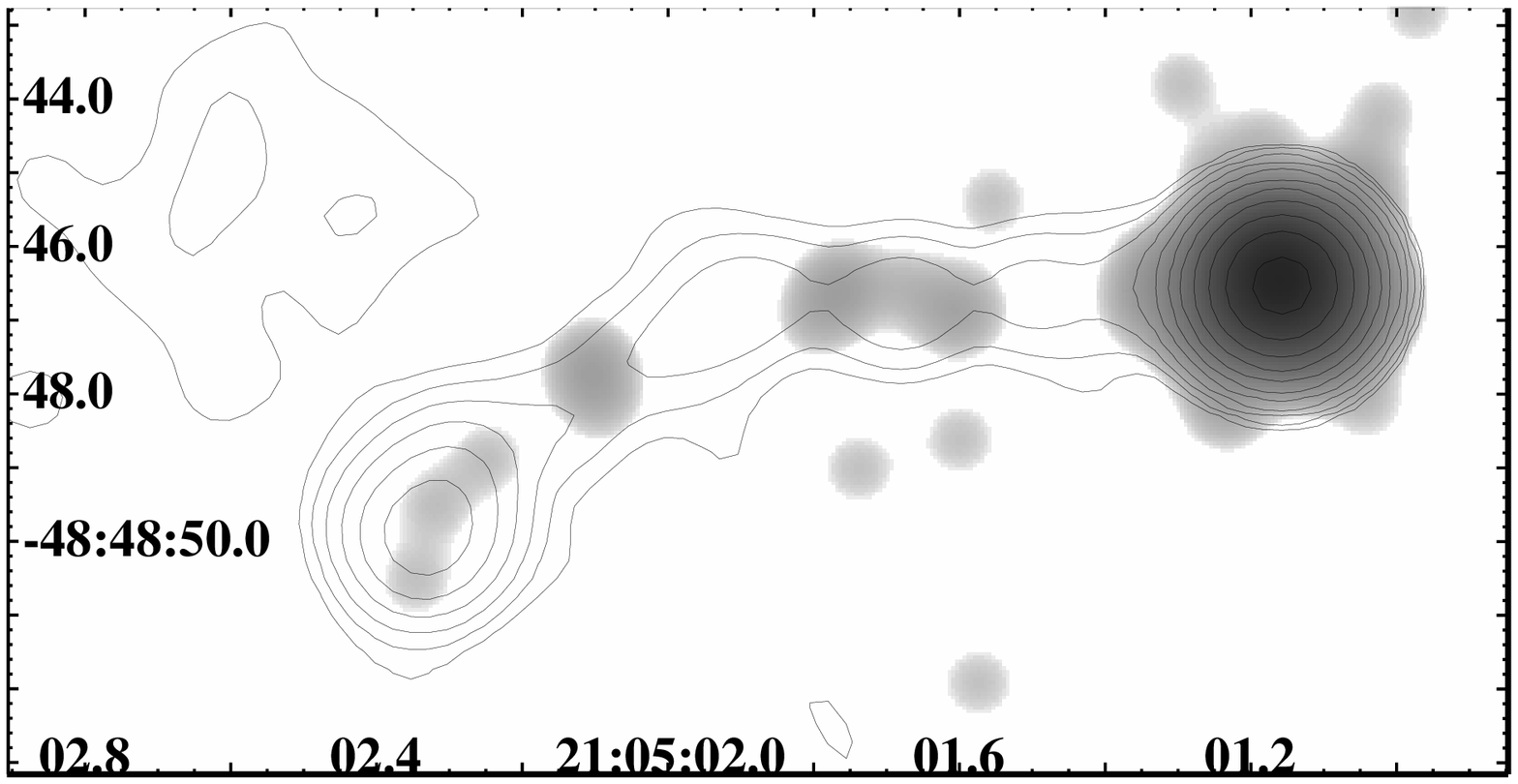}
\end{minipage}
\caption{\label{figure1}X-ray images with radio contour overlays for
  0920--397, 1202--262, 1030--357 \& 2101--490 (left to right). 
  0920 and 1202 are representative of most jet systems:
  the X-rays match the radio contours, 
  fading rapidly after the sharp bend $5.6''$ NNW of the core of 1202.
  Uniquely, the X-rays in 1030 remain strong through a sequence of
  sharp bends.
  2101 is unusual in that the X-ray knots lie between the radio peaks.
}
\end{figure}

X-ray jets are detected in 12/20 sources;
a higher 
rate amongst the (radio) brighter jets suggests 
deeper X-ray observations would yield more detections.
X-ray jets are one sided, with peaks usually coincident with radio
knots up to the first sharp bend
(Fig.~\ref{figure1}).
Low optical flux limits in five systems indicate that the
synchrotron continua cut off below $\nu \sim \! 10^{14}$~Hz, suggesting
that the X-ray flux is dominated by a different process.
IC-CMB models\cite{tavecchio} fit best,
suggest
bulk Lorentz factors $\Gamma \! \sim \! 3$--10 and magnetic fields 
$B_\mathrm{eq} \! \sim \!
10^{-5}$~Gauss, with jets directed within $\sim \! 20^\circ$ of our line of
sight\cite{marshall04,schwartz03}.


%
%
%
%


\begin{thebibliography}{0}
\bibitem{marshall04} H. Marshall {\it et al.}, {\it ApJS, submitted};
  see http://space.mit.edu/\linebreak[0]$\sim$jonathan/\linebreak[0]jets.

\bibitem{gelbord04} J. Gelbord {\it et al.}, {\it in prep.}

\bibitem{tavecchio} F. Tavecchio {\it et al.}, {\it ApJ} {\bf 544},
  L23 (2000).

\bibitem{schwartz03} D. Schwartz {\it et al.}, {\it New Astron. Rev.}
  {\bf 47}, 461 (2003).

\end{thebibliography}
\end{document}